# Impedance-Guided Programmable Transmission of Localized Deformation in Modular Soft Metamaterials

## Authors Information

Weiyun Xu,[1]* Daewon Hong,[2] Zhi Zhao,[1] Rahul Dev Kundu,[1] Xiaojia Shelly Zhang[1,2,3]*

[1] Department of Civil and Environmental Engineering, University of Illinois Urbana-Champaign, Urbana, IL 61801, USA.

[2] Department of Mechanical Engineering, University of Illinois Urbana-Champaign, Urbana, IL 61801, USA.

[3] National Center for Supercomputing Applications, Urbana, IL 61801, USA.

* Correspondence to: Weiyun Xu (weiyunxu@illinois.edu) and Xiaojia Shelly Zhang (zhangxs@illinois.edu).

## Abstract

Soft mechanical metamaterials provide a promising platform for applications in robotics, biomedical devices, and flexible electronics. The localized mechanical responses induced by nonuniform excitation are ubiquitous in soft materials, yet their controlled transmission across modular metamaterial assemblies remains largely overlooked in metamaterial design, which critically constrains nontrivial functionalities that require end-to-end and long-range deformation transmission. Here, we introduce an impedance-guided design framework that enables programmable transmission of localized deformation in modular soft metamaterials, achieving behaviors unattainable by intuitive design. By establishing an equivalent nonlinear spring model considering position-dependent interactions and integrating the concept of mechanical impedance within soft metamaterials, we regulate assembly-level transmission solely through unit-cell topology optimization. The resulting framework enables effective synthesis of module families, allowing both homogeneous and heterogeneous assemblies to be custom-built with markedly enhanced transmission characteristics. Leveraging the highly combinatorial and extensible design space, we physically realize diverse on-demand displacement manipulation architectures, including obstacle-bypassing modular soft-metamaterial assemblies, defect-tolerant soft gripping, and embodied signal processing. Beyond deformation programming, the reconfigurability and reassemblability of these soft modules can embed electric logic signals, enabling energy-efficient and low-latency information processing through compliant-switch-controlled mechanical LED displays and wearable finger-motion-sensing controllers. Our method provides fundamental insights into localized deformation transmission in modular soft metamaterials and establishes a scalable route toward embodied-intelligence material systems, particularly for soft-metamaterial-centric actuation, sensing, and collective computing.

## Introduction

Metamaterials have emerged as a powerful platform for systems with physical intelligence, owing to their ability to achieve exceptional and programmable mechanical responses through the rational design of microstructures and constituent materials [1,2]. In particular, soft metamaterials, which employ compliant matrices such as elastomeric polyurethane (EPU) [3], polydimethylsiloxane (PDMS) [4], liquid crystal elastomers (LCEs) [5], and hydrogels [6], are distinguished by their large deformability and resilient recoverability. Driven by the combination of soft matter and architected geometry, extensive efforts have been devoted to nonlinear mechanical behavior [7] and multiphysics integration [8] in soft metamaterials, enabling applications in soft robotics [9], electronic skin [10], human-computer interaction [11], and material-level information processing [12]. Many of these emerging functionalities would particularly benefit from modular, scalable, and reconfigurable architectures [13-15], which are essential for large-scale and adaptive material systems. Despite these advances, most existing studies focus on the global nonlinear response of monolithic or weakly modular structures, such as a small number of periodic unit cells [16,17] under relatively uniform loading conditions [18]. By contrast, localized mechanical responses induced by nonuniform excitation, local deformation, or biased boundary conditions, which are often relevant in practice, remain much less explored in the static regime. This limitation is especially consequential in soft materials, which naturally convert local mechanical inputs into distributed strain energy, thus attenuating the transmitted deformation and force over distance. Such attenuation fundamentally constrains the modularity, scalability, and reconfigurability of soft metamaterial systems.

The fundamental origin of this issue can be illustrated by a simple hyperelastic body subjected to a localized input (Fig. 1A). When a bulk elastomer is displaced at one boundary location while anchored at four corners, the induced stress and displacement fields remain strongly localized, and both the output displacement and the reaction force at the input decay rapidly with distance from the excited region. This behavior is consistent with Saint-Venant's principle [19,20] and arises broadly in structures subjected to spatially localized or discontinuous loading. A similar phenomenon persists in cascaded soft metamaterial assemblies. For a chain composed of $N$ connected unit cells (Fig. 1B), a prescribed input displacement $D_{in}$ applied to the first cell defines an equilibrium working point for the entire assembly, including the input reaction force $F_{in}$, the internal strain distribution, and the terminal output displacement $D_{out}$. In general, $D_{out}$ decays rapidly with the number of cells, analogous to the spatial attenuation observed in bulk soft solids.

Although attenuation of local deformation is desirable in applications such as vibration isolation, cushioning, and energy absorption [21-23], many emerging soft material systems instead require sustained end-to-end transmission of mechanical signals, forces, or displacements, as in soft actuation [24], compliant mechanisms [25], and deployable structures [26]. For example, Byun et al. used cascaded soft machines to process mechanical wave signals and showed that dissipation can suppress propagation between neighboring units [27]. The varying mechanical responses of auxetic metamaterials under heterogeneous driven conditions were characterized as anomalous size effects by Coulais et al. [28], which emphasized the critical role of mechanism-based metamaterials for preserving functionality from elastic hybridization in the large-scale limit. In purely passive soft material systems, and in the absence of additional or prestored energy such as bistability, attenuation of localized deformation appears difficult to avoid. Developing strategies to mitigate this attenuation thus represents a central challenge in the design of modular soft metamaterials capable of long-range deformation transmission.

Conventional design approaches are not well-suited to this problem. Heuristic strategies, including intuitive and bio-inspired designs [29-31], heavily rely on physical intuition, which may become limited for long-range deformation transmission under localized loading governed by complex assembly-level interactions. Computational methods, including topology optimization [5] and machine learning [32-34], offer much greater design flexibility in geometry generation, but their direct application could become increasingly impractical when the target response depends on the coupled behavior of many interacting unit cells. Directly simulating large assemblies of hyperelastic metamaterials is computationally expensive and can be numerically challenging, because it requires high-resolution, constitutive descriptions beyond linear elasticity [35], and nonlinear numerical frameworks such as the finite element method [36] and the material point method [37]. At the same time, the apparent simplicity of an individual unit cell is difficult to exploit, because its effective behavior depends strongly on its position within the assembly, the downstream mechanical resistance, and the global configuration of the structure. In these localized deformation transmission problems, periodic boundary conditions are generally inapplicable.

Here, we address this gap by systematically investigating the attenuation of localized deformation and introducing an impedance-guided computational framework for programmable localized deformation transmission in modular soft metamaterials. Rather than optimizing isolated unit cells, the proposed framework explicitly incorporates assembly-level transmission into the unit-cell design process. We develop modular soft metamaterial unit cells that exhibit programmable deformation manipulation under unilateral input when assembled and reassembled into diverse architectures. Building on a nonlinear effective spring-stiffness description of an individual unit cell, we extend the concept of mechanical impedance from dynamic systems to the nonlinear quasi-static incremental regime by representing the downstream subassembly as an input impedance seen at the connected port (Fig. 1C). On this basis, we establish an impedance-guided design approach that captures assembly-level behavior through transmission-aware unit-cell topology optimization. With strong reconfigurability and extensibility, the resulting designs markedly delay attenuation of localized deformation in cascaded soft assemblies and establish two-port and four-port module families that provide a wide design space of the corresponding deformation-transmission modes (Fig. 1D). We further show that these optimized behaviors are unattainable through intuitive design alone and that the resulting modules enable a diverse set of programmable functionalities in modular assemblies, including alternating compliant switching, on-demand displacement routing, defect-tolerant soft gripping, and wearable finger-motion sensing and control (Fig. 1E). Using experimentally characterized 3D-printed TPU, we validate the predicted behaviors and demonstrate the physical realizability of the proposed designs. More broadly, this work clarifies the mechanism governing localized displacement transmission and establishes it as a new design component for modular soft metamaterials, provides an efficient framework for understanding assembly-level mechanical interactions, and offers practical design principles for intelligent material systems in applications ranging from soft robotic actuation to self-sensing metastructures and modular robotic locomotion.

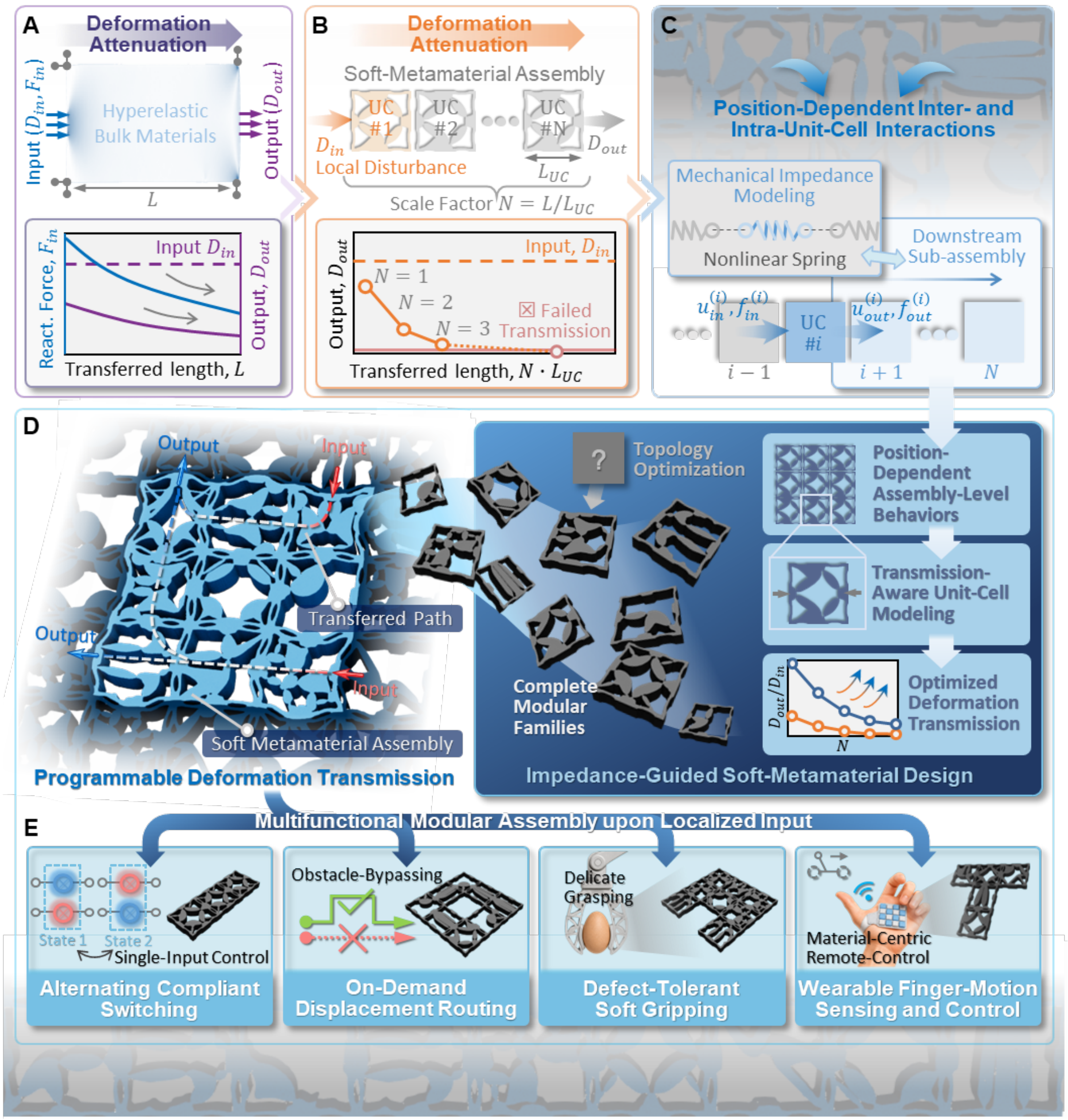

**Fig. 1. Impedance-guided programmable deformation transmission in modular soft metamaterials.** **(A)** Localized deformation of a hyperelastic bulk material under a prescribed nonuniform input $(D_{in}, F_{in})$, illustrated by the von Mises stress distribution. The attenuation of the input reaction force and the output displacement along the transferred length is plotted below. **(B)** Attenuation of deformation transmission in a cascaded assembly of soft unit cells, where a localized disturbance is applied to the first unit cell and the output is measured at the $N$th unit cell. The corresponding decay of the output displacement is plotted below. **(C)** Assembly-level position-dependent interactions among unit cells modelled through an equivalent mechanical impedance, in which the downstream subassembly is represented as an equivalent nonlinear spring. **(D)** Impedance-guided design of modular soft-metamaterial families for programmable deformation transmission. **(E)** Representative multifunctional modular soft metamaterials, including an array of alternating compliant switches, on-demand displacement routing, a defect-tolerant soft gripper, and a wearable finger-motion sensing controller.

# Results

## Analysis of the Deformation Transmission in Soft Metamaterials

### *Deformation attenuation in cascaded modular metamaterials*

To investigate the localized deformation transmission in a modular assembly of soft mechanical metamaterials, we consider two representative types of compliant displacement transformers as illustrated in Fig. 2. In the first case, the deformation within a unit cell is transferred along the imposed load direction, which can be achieved by many heuristic unit-cell "transmitter" designs, such as cross-shaped lattice (CL) and bar-shaped lattice (BL). In contrast, the second case enabled by a diamond-shaped lattice (DL) or diamond-perforated solid (DS) unit cell can reverse the direction of input, which functions as a displacement "inverter" compliant mechanism. The specific dimensions of each design are shown in Supplementary Fig. S1. Without loss of generality, we set the unit cell size as $L \times L = 50\text{ mm} \times 50\text{ mm}$, and impose an $x$-directional Dirichlet boundary condition at the left-middle region of the first unit cell with inward $D_{in} = 0.1L$ as the working point of the cascaded metamaterial assembly. Each unit cell is anchored at the four corners to facilitate its modularity. Therefore, the transmission efficiency (TE) in the localized distribution of deformation can be evaluated by the normalized average outlet displacement (output-to-input ratio) at the right-middle region, i.e., $\text{TE} = |D_{out}/D_{in}|$, and the corresponding average reaction force at the input, $F_{in}$. Their variation with respect to the number of cascaded unit cells (i.e., scale factor) $N$, based on the four designs, is presented in Figs. 2A and 2B, respectively.

Throughout this work, the mechanical behaviors of soft metamaterials considering large deformation and self-contact are simulated numerically by nonlinear finite element analysis (FEA) integrated with the Lopez-Pamies (LP) hyperelastic continuum model [38] and the third medium method [39] (see Materials and Methods below). The resulting FEA in Fig. 2A demonstrates a rapid decay of the locally-actuated deformation for both the displacement transmitter and the inverter as $N$ increases. With $|D_{out}/D_{in}| = 1$ representing lossless transmission, both cases already fall below this threshold at $N = 1$. Intuitively, the transmitter designs preserve the sign of displacement similar to that of the bulk hyperelastic materials. While the inverter designs show a unique sign-alternating transmission along the assembled unit cells, the attenuation is much more severe, with $|D_{out}/D_{in}| \leq 0.1$ since $N = 3$.

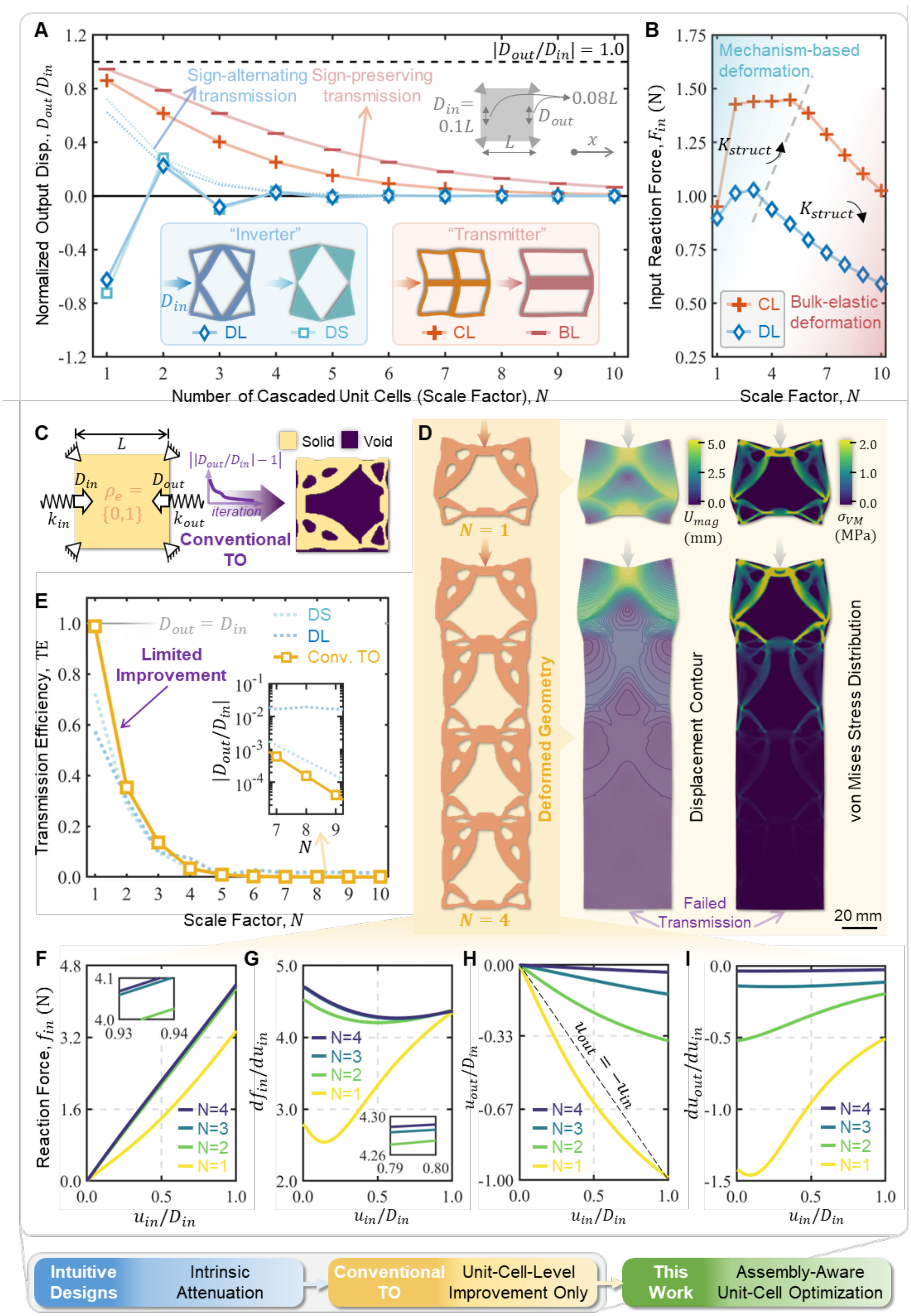

**Fig. 2. Analysis of deformation transmission shows localized deformation attenuates rapidly in cascaded soft metamaterials.**

**(A)** Transmission efficiency (i.e., normalized output displacement) as a function of assembly size $N$, for assemblies composed of the cross-shaped lattice (CL), bar-shaped lattice (BL), diamond-shaped lattice (DL), and diamond-perforated solid (DS) unit cells, respectively. The transmitter-type designs (CL and BL) exhibit sign-preserving transmission, whereas the inverter-type designs (DL and DS) show sign-alternating transmission. **(B)** Corresponding variation of the input reaction force with increasing assembly size $N$. **(C)** Conventional topology optimization for compliant mechanism synthesis and the resulting optimized modular unit cell. **(D)** Deformed geometry, displacement-magnitude contour, and von Mises stress distribution of the conventional TO design at $N = 1$ and $N = 4$. **(E)** Comparison of transmission attenuation among the DS, DL, and conventional TO designs. **(F)** Normalized input force-displacement curves of the conventional TO design, together with the corresponding **(G)** incremental input stiffness. **(H)** Normalized output-input displacement curves for the conventional TO design, together with the corresponding **(I)** incremental displacement transmission ratio.

The non-monotonic trend of reaction force (Fig. 2B) further reveals a mode switch of the architected materials from the mechanism-based deformation to the bulk-elastic deformation. The mechanism-based deformation is the soft mode of compliant mechanism, where rotation, bending, and buckling dominate to prevent fluctuation between adjacent unit cells [28]. Initially, the soft-mode is favored to satisfy the geometry-compatible interaction with an increased effective stiffness of the assembly, indicating effective transmission. As $N$ increases, the whole assembly has more space to re-distribute the deformation gradient, localizing the soft mode and making the transmission analogous to the boundary effect of bulk materials. The input displacement can hardly actuate the whole cascaded array and the force can be barely transferred since the bulk-elastic mode is dominant. It suggests that, as the effective stiffness of the soft metamaterials changes dynamically, optimization is required to enlarge the affected region of the local disturbance based on the interactions intra- and inter- unit cells.

While topology optimization (TO) has proven effective for obtaining optimized material and structural properties based on gradient information from the objective function and constraints, we found that a conventional TO framework for compliant mechanism synthesis has limited capability to mitigate the attenuation described above. For demonstration, we implemented TO on the inverter case of the soft metamaterials, which is more challenging with alternating deformation directions and tensile-compressive stress distributions. The objective is to minimize the transmission loss as quantified by $||D_{out}/D_{in}| - 1|$, with the problem setup and the optimized modular unit cell geometry presented in Fig. 2C (see Materials and Methods below for details). As shown by the deformed geometries and displacement contours (Fig. 2D), the conventional TO design achieves $\mathrm{TE} = 1$ for a single unit cell. However, transmission nearly vanishes at $N = 4$, where the output displacement falls below $1\%$ of the input. The von Mises stress distribution in Fig. 2D further implies the localization of transferred force. Compared with the two intuitive inverter designs (DL and DS), the conventional TO design shows an increased output displacement only at $N = 1$, followed by a much steeper decay as additional unit cells are assembled in a chain (Fig. 2E). To further elucidate the nonlinear responses associated with different assembly states ($N = 1$ to $N = 4$), Figs. 2F-I present the normalized input force-displacement ($f_{in} - \frac{u_{in}}{D_{in}}$) curves, the normalized output-input displacement ($\frac{u_{out}}{D_{in}} - \frac{u_{in}}{D_{in}}$) curves, and their corresponding first derivatives with respect to the input displacement, for the conventional TO design. The slope of the input force-displacement curve, $df_{in}/du_{in}$, is referred to as the incremental input stiffness, while $du_{out}/du_{in}$ is termed the incremental displacement transmission ratio. The curves of the DL design are shown in Supplementary Fig. S2 for comparison. These results suggest that interactions in soft metamaterials vary with both the assembly configuration and the loading state along the path from the initial to the working point. Optimizing an individual unit cell along cannot capture the spatially varying external resistance imposed by the cascaded assembly. Moreover, the intrinsic nonlinearity of soft matter must be characterized along the entire loading path, rather than only at the working point. At the same time, nonlinear FEA scales as $O(n^d)$ with the number of unit cells, making direct simulation of the full assembly increasingly impractical as $N$ becomes large (Supplementary Fig. S3).

*Effective spring-stiffness modeling of modular unit cell*

Guided by the transmission analysis and the underlying intra- and inter- cell interactions, we establish a general equivalent model for the hyperelastic unit cell based on a nonlinear spring network and quantify the assembly by the map of mechanical impedance. Leveraging the modularity of the unit cell, the model captures the assembly-level mechanical response in an analytically tractable manner and can be readily incorporated into the TO framework.

Specifically, the intra-cell input-output relation of a soft unit cell can be equivalently described by an effective elastic spring, $k_{LR}$, which connects a left input region (port L) and a right output region (port R) (Fig. 3A). Additionally, each port is connected to the fixed reference defined by the corner supports through a separate grounded spring, $k_L$ or $k_R$, which represents the effective self-stiffness arising from deformation of the material paths linking the respective port to the fixed boundaries. Therefore, the unit cell is characterized by an effective tangential stiffness matrix $\boldsymbol{K}$ (see Materials and Methods; Supplementary Text S1) at a prescribed working point $D_{in}$. The stiffness components are extracted using a Schur complement-based procedure, through an auxiliary linearization of the residual $\boldsymbol{R}$ obtained from the nonlinear finite element analysis (see Supplementary Text S2). To account for the external load applied at port R, the $K_{22}$ component is modified to $K_{22} + k_{out}$, where $k_{out}$ denotes the nonlinear stiffness associated with the external resistance. For a natural boundary, $k_{out} = 0$.

Having established the equivalent description at the single-cell level (Fig. 3B), we next extend the analysis to the interactions between adjacent unit cells in a cascaded assembly with a lossless connection. For an $\bar{N} \times 1$ array of two-port unit cells, each cell is no longer terminated by a prescribed boundary condition, but by the effective mechanical resistance of the downstream cells (Fig. 3C). It provides a natural static analogue of mechanical impedance in the assembled system, where the input impedance of the $i+1$ to $\bar{N}$ downstream subassembly can be defined as the incremental force-displacement relation at the input port of cell $i+1$ (Fig. 3D). Therefore, the intra-cell interaction behavior with a match of mechanical impedance leads to

$$k_{out}^{(i)} = Z_{in}^{(i+1)}. \tag{1}$$

Accordingly, using a recursive formula of Riccati-form, we define the modular transmission ratio, $r^{(i)}$, and the input impedance for the $i$-th unit, $Z_{in}^{(i)}$ (see details in Materials and Methods). The total displacement gain from the first input to the $\bar{N}$-th output, $r_{tot}^{(\bar{N})}$, is also defined, indicating the deformation transmission's dependence on the length of the cascaded array. While evaluating the transmission efficiency requires multiple control points along the loading history and integration of the total displacement gain, the model provides an effective and accurate description of transmission, as validated by comparing the directly computed FEA responses of the full assemblies for the four intuitive designs with the values predicted from the proposed method using only unit-cell-level analysis with 3 control points (Supplementary Table S1). With the modular transmission ratio $|r^{(i)}|$ typically smaller than unity in soft materials, the gain also suggests an exponential decay, which is evidenced by Fig. 1A.

Importantly, the proposed nonlinear spring model and impedance-based formulation are sufficiently general to describe both displacement inverters ($\mathrm{sign}(r^{(i)}) = -1$) and transmitters ($\mathrm{sign}(r^{(i)}) = +1$). They are not restricted to collinear two-port configurations, but extend directly to arbitrary port placements, to both periodic and nonperiodic assemblies, and to three-dimensional geometries (Fig. 3E). For instance, a four-port representation can be formulated by taking the four edges of a square unit cell as ports, yielding a corresponding $4 \times 4$ effective tangential stiffness matrix $\boldsymbol{K}$. Additionally, a periodic configuration composed of identical unit cells gives rise to a homogeneous assembly, in which all unit cells share the same stiffness matrix, whereas a nonperiodic configuration results in a heterogeneous assembly with cell-dependent $\boldsymbol{K}^{(i)}$. Given the homogeneous assembly and a free end at the final output ($k_{out}^{(\bar{N})} = 0$), the explicit forms of the modular transmission ratio $r^{(i)}$, the input impedance $Z_{in}^{(i)}$, and the total displacement gain $r_{tot}^{(\bar{N})}$ from $\bar{N} = 1$ to $\bar{N} = 3$ are derived in Supplementary Table S2.

Moreover, for a homogeneous assembly, the impedance recursion (Eq. 9) converges to a fixed point given $|r^{(i)}| < 1$, yielding a stabilized input impedance, $Z_{in}^{*}$, and a corresponding stabilized modular transmission ratio, $r^{*}$ (see Supplementary Text S3). These quantities characterize the asymptotic transmission behavior of a half-infinite cascaded chain ($\bar{N} \to +\infty$). To quantify the resulting attenuation, we define a characteristic attenuation length

$$l(D_{in}) = \frac{1}{|\ln|r^{*}(D_{in})||}, \tag{2}$$

such that, at $u_{in} = D_{in}$, the incremental displacement gain decays to $1/e$ of its initial value after propagating across $l$ unit cells. Notably, this characteristic attenuation length also serves as a useful indicator of the resistance of a unit cell to input transmission, and thus remains informative for heterogeneous assemblies.

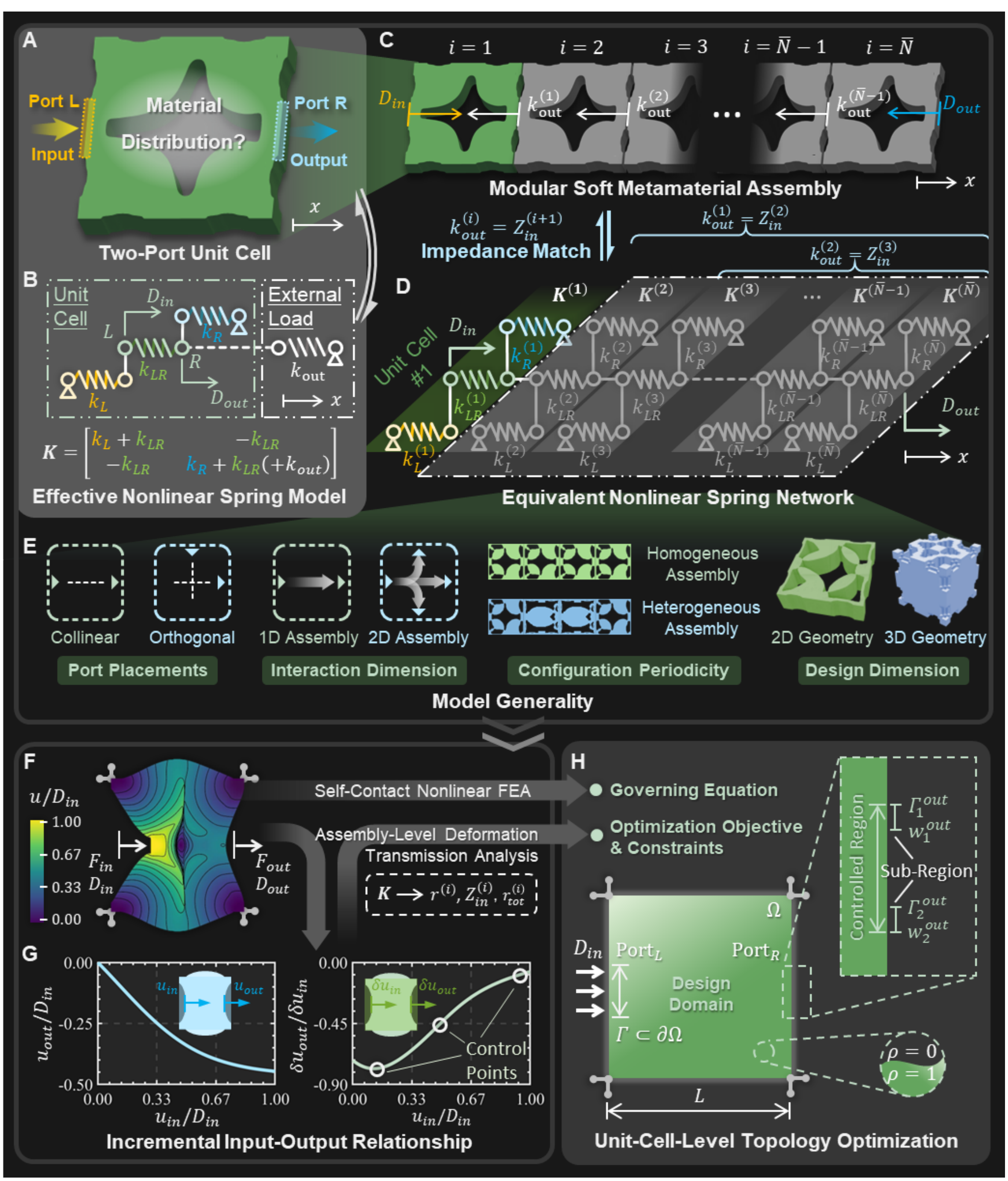

**Fig. 3. Impedance-guided framework for assembly-aware topology optimization.**
**(A)** Schematic of a modular unit cell with collinear input port L and output port R. **(B)** Equivalent one-dimensional nonlinear spring model of the two-port unit cell. **(C)** Cascaded soft-metamaterial assembly with position-dependent external output loading, $k_{out}^{(i)}$. **(D)** Spring-network representation of inter-cell interactions based on impedance matching. **(E)** Generality of the proposed framework, including different port placements, interaction dimensionalities (two-port versus four-port), assembly periodicities (homogeneous versus heterogeneous), and geometric dimensionalities. **(F)** Nonlinear finite element analysis (FEA) accounting for self-contact. **(G)** Normalized output-input displacement curves and the corresponding incremental displacement transmission ratios. Based on the effective stiffness $\boldsymbol{K}$, the assembly-level deformation transmission is characterized by the modular transmission ratio $r^{(i)}$, the input impedance $Z_{in}^{(i)}$, and the total displacement gain $r_{tot}^{(i)}$. **(H)** Problem setup for the impedance-guided unit-cell-level topology optimization.

## Impedance-Guided Module Synthesis for Soft Metamaterials

*TO design framework for delaying the transmission attenuation*

Motivated by the intrinsic attenuation of displacement transmission in soft materials, as revealed by the total displacement gain $r_{tot}^{(\bar{N})}$, we proposed an impedance-guided design framework for architected soft metamaterials that mitigates this decay and supports scalable modular applications. Built on compliant-mechanism-synthesis TO (Fig. 2C) and nonlinear finite element modeling with self-contact (Fig. 3F), this framework uses the mechanical impedance underlying the incremental input-output relationship (Fig. 3G) to guide the design process, thereby enabling efficient unit-cell-level optimization while accounting for assembly-level deformation transmission (see Materials and Methods; Fig. 3H).

To preserve transmission over long chains in general, we first propose a global optimization strategy which maximizes the characteristic attenuation length. Following the foregoing analysis, the loading path is discretized by $M$ control points, $u^m \in (0, D_{in}]$ $(m = 1, \ \dots\ , M)$, as shown in Fig. 3G. The optimization problem is then written as

$$\begin{aligned} \max_{\boldsymbol{\rho}} \quad & \min_{m=1,\dots,M} \{1/|\ln|r^*(u^m)||\} - \eta V(\boldsymbol{\rho}) \\ \text{s.t.} \quad & \sum_{m=1}^{M} \omega_m \left| r_{tot}^{(i)}(u^m) \right| \geq \alpha \sum_{m=1}^{M} \omega_m \\ & \left( K_{11} + 2K_{21} r_{tot}^{(i)} + K_{22} {r_{tot}^{(i)}}^2 \right) / 2 \leq e_{max}, \ \forall m \\ & K_{11} \geq \tau_A, \ K_{22} \geq \tau_B, \ \det \boldsymbol{K} \geq \delta_{min}, \ \forall m \\ & V(\boldsymbol{\rho}) - V^* \leq 0 \\ & \sigma_{VM}^M(\boldsymbol{\rho}) - \sigma^* \leq 0 \\ & \boldsymbol{0} \leq \boldsymbol{\rho} = \{\bar{\rho}_e\} \leq \boldsymbol{1}, \ e = 1, \ \dots, \ N_e \end{aligned} \tag{3}$$

with $\boldsymbol{R}(\boldsymbol{\rho}, \boldsymbol{u^m}) = \boldsymbol{0}$ at $u_{in} = u^m$ on $\Gamma \subset \partial\Omega$. Here, the penalty term $-\eta V(\boldsymbol{\rho})$ is introduced to suppress redundant material usage. The Riemann sum of the total displacement gain $r_{tot}^{(i)}(u^m)$, evaluated with trapezoidal weight $\omega_m$, is constrained to exceed a prescribed lower bound $\alpha$, thereby enforcing $|D_{out}/D_{in}| \approx 1$. An energy constraint, $e_{max}$, is imposed to limit elastic energy storage within the soft material, thereby preventing excessive input work from being dissipated through localized deformation within an individual unit cell (see Supplementary Text S4). The thresholds $\tau_A$, $\tau_B$, and $\delta_{min}$ enforce the positive-definiteness requirements of the effective stiffness. $V^*$ and $\sigma^*$ denote the prescribed limits for volume fraction and von Mises stress, respectively, to reduced material use, ensure connectivity, and mitigate overly thin members.

Notably, the global optimization strategy does not explicitly account for the actual assembly-specific interactions during optimization. Instead, each unit cell is designed to improve general transmission characteristics, which substantially simplifies the optimization procedure while remaining effective for capturing the transmission capability of nonperiodic assemblies, even though the underlying metrics are derived from periodic configurations. In contrast, for cases that fully account for cell-to-cell variations in transmission behavior, including external loading and transmission direction, we develop a recursive optimization strategy to generate case-specific unit-cell designs one by one. In this process, the optimized unit cells from $\bar{N}$ to $i+1$ collectively act as the external load attached to the $i$th cell, and the constrained total displacement gain $r_{tot}^{(i)}$ is evaluated according to Eqs. 8-10. Starting from $i=\bar{N}$ and proceeding to $i=1$, each unit cell in the chain undergoes one round of topology optimization referring to Eq. 3. Consequently, $\bar{N}$ optimization subproblems are formulated and solved during this procedure, generating a specific assembly with optimized transmission performance.

Integrated with the global and recursive optimization strategies, the overall design workflow is summarized in Supplementary Fig. S4. Taken together, the proposed impedance-guided module synthesis simultaneously accounts for unit-cell scalability, strain energy conversion, displacement-direction regulation, and effective-stiffness stability, thereby establishing a general design framework for modular soft-metamaterial assemblies. In the following subsections, we demonstrate this capacity through periodic collinear designs, two-port module-family generation, and four-port module-family generation; the corresponding optimization parameters are listed in Supplementary Table S3.

*Periodic collinear designs*

Using the global optimization strategy established for homogeneous assemblies, we obtained optimized periodic designs with substantially reduced deformation attenuation. As a representative example, we consider designing the collinear two-port displacement inverter (hereafter termed the collinear inverter, CI), whose transmission behaviors can be tailored by constraining different total displacement gain $r_{tot}^{(i)}$ in Eq. 3. To emphasize the intrinsic origin of deformation attenuation in soft metamaterials, we first plot the heat maps of the gain for $i=1\text{-}3$ at working point as functions of the normalized stiffness components $K_{21}/K_{11}$ and $K_{22}/K_{11}$ (Fig. 4A and Supplementary Fig. S5). It shows that, as the number of unit cells increases, the potential material parameters required to achieve lossless transmission ($r_{tot}^{(i)} \geq 1.0$) are rapidly driven toward the boundary of the positive-definite domain ($\det \boldsymbol{K} \geq 0$), which becomes increasingly narrow and extreme. This trend highlights the difficulty of mitigating it through intuitive design alone. To address this challenge, we solved three variants of the optimization problem by separately constraining $r_{tot}^{(1)}$, $r_{tot}^{(2)}$, and $r_{tot}^{(3)}$ (Supplementary Table S2), which led to three converged designs (Designs 1-3), denoted as CI-1, CI-2, and CI-3, respectively. The corresponding evolution paths of $r_{tot}^{(1)}$ from the initial guess (IG 1) to the final designs are shown in Fig. 4A. We also investigated three rationally designed initial guesses (IGs 1-3) with distinct effective stiffness matrices $\boldsymbol{K}$ for each constrained case (Supplementary Fig. S5), which notably exhibits a clustered pattern of the optimized solutions according to the constrained $r_{tot}^{(i)}$, with each group converging toward its corresponding feasible property domain. This result suggests that the optimized design is primarily governed by the target transmission requirement rather than the artificial initialization. Given IG 1, the resulting geometries and iteration histories (iteration 0-1000) are provided in Supplementary Fig. S6.

The impedance-based TO designs show markedly delayed attenuation, as reflected by the decay of the transmission efficiency in Fig. 4B. Compared with both IG 1 and the diamond-perforated solid (DS, Fig. 2A), the synthesized CI Designs 1-3 all transmit observable displacements to more distant unit cells along the assembly direction while preserving the displacement-inverting functionality. The transmitted displacements at $\overline{N} = 1 \sim 3$ were further validated experimentally using 3D-printed TPU samples of CI Designs 1-3 (Figs. 4B and 4C), which agree well with the numerical predictions (see Materials and Methods). Consistent with the optimized characteristic attenuation length $l$ (Supplementary Fig. S6), these results indicate that the mechanical response of the soft metamaterial can be effectively regulated by representing the unit-cell load-bearing behavior as the interaction in a nonlinear spring network. The enhanced transmission of local deformation in the optimized periodic assemblies is further evidenced by the displacement contours at $\overline{N} = 6$ (Fig. 4D): whereas the DS benchmark design exhibits nearly failed transmission, with $|D_{out}/D_{in}| < 0.5\%$, CI Designs 1-3 achieve at least 12-fold larger output displacements over a transfer distance reaching 60 times the input magnitude.

Although the three designs exhibit distinct material distributions within the unit cell, they share an emerging mechanized morphology characterized by hinge-like connections between thick rods. This feature promotes the soft deformation mode discussed in Fig. 2B and reduces elastic energy storage in the soft metamaterial. The deformation pattern and corresponding von Mises stress contour of CI-2 at $\overline{N} = 3$ (Fig. 4E), which are validated by the actuation experiment (Fig. 4F), further reveal the transmitted force path and the stress concentration near the hinge regions, consistent with the mechanism-based deformation mode. Meanwhile, we validated the design framework by confirming that the results are insensitive to displacement or force control input boundary condition, and the generated geometries exhibit properties consistent with the quantified stiffness matrix $\boldsymbol{K}$. In particular, the resulting mechanical responses preserve tension-compression symmetry and reciprocity, while retaining input-output asymmetry (Supplementary Fig. S7).

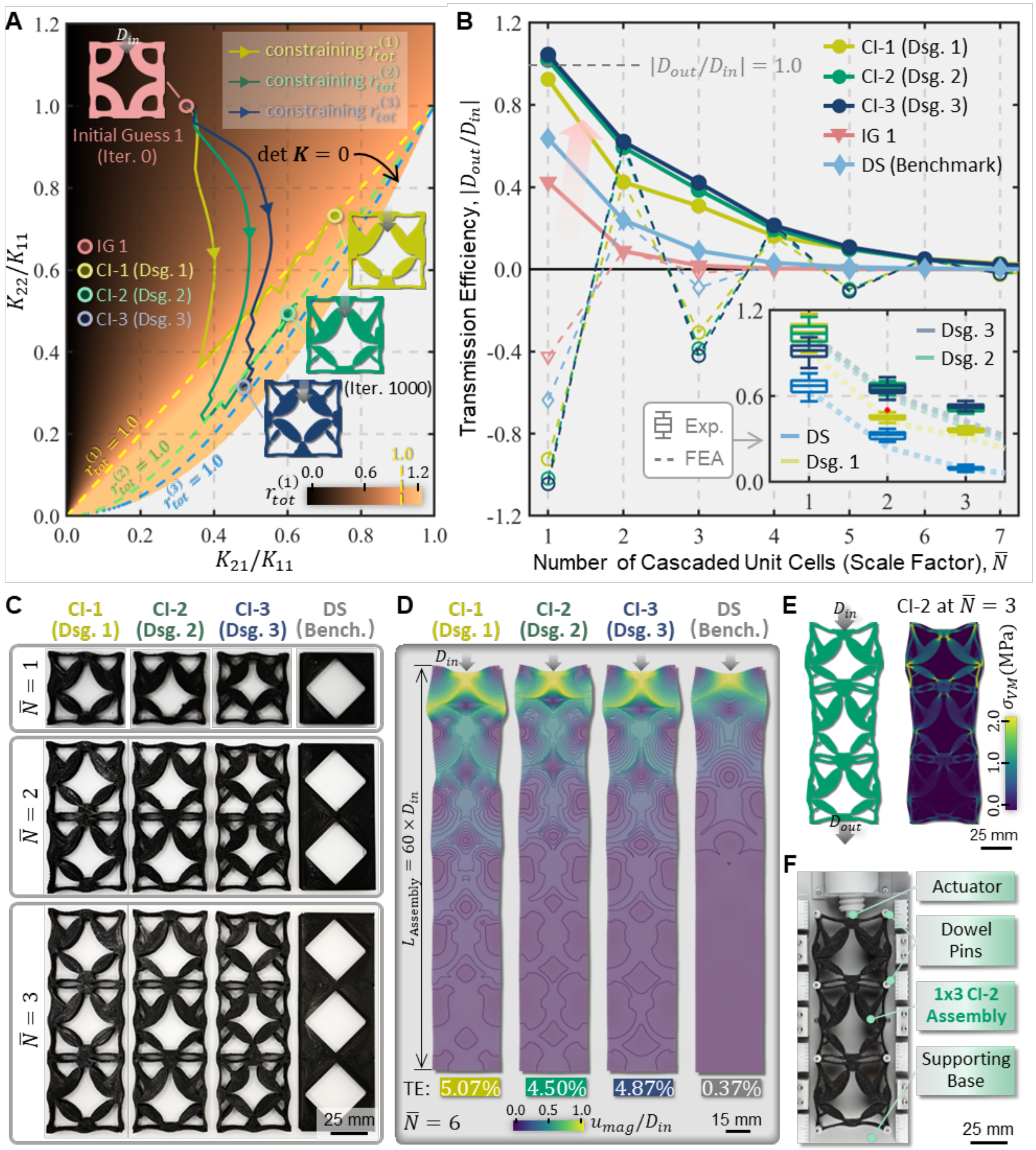

**Fig. 4. Impedance-guided design of periodic collinear soft-metamaterial unit cells with enhanced deformation transmission.**

**(A)** Heat map of the total displacement gain $r_{tot}^{(1)}$ at the working point as a function of the normalized stiffness components $K_{21}/K_{11}$ and $K_{22}/K_{11}$. The boundary is defined by the determinant of the stiffness matrix, $\det \boldsymbol{K} = 0$. The evolution paths trace the change in $r_{tot}^{(1)}$ from iteration 0 (Initial Guess 1) to iteration 1000 for three optimization problems, in which $r_{tot}^{(1)}$, $r_{tot}^{(2)}$, and $r_{tot}^{(3)}$ are constrained separately, yielding Designs 1-3, respectively. The yellow, green, and blue dash lines denote $r_{tot}^{(1)} = 1$, $r_{tot}^{(2)} = 1$, and $r_{tot}^{(3)} = 1$, respectively. **(B)** Transmission efficiency as a function of assembly size, comparing the numerical responses of CI Designs 1-3, Initial Guess 1, and the diamond-perforated solid (DS) benchmark. The inset compares the numerical and experimental results of Designs 1-3 and the DS benchmark for $\bar{N} = 1 - 3$. **(C)** Fabricated TPU samples of Designs 1-3 and the DS benchmark (see Materials and Methods). **(D)** Displacement-magnitude contours of Designs 1-3 and the DS benchmark at $\bar{N} = 6$. For $D_{in} = L_{Assembly}/60 = 5$ mm, the transmission efficiency, TE $= |D_{out}/D_{in}|$, is $5.07\%$, $4.50\%$, $4.87\%$, and $0.37\%$, respectively. **(E)** Deformed geometry and corresponding von Mises stress distribution of CI Design 2 at $\bar{N} = 3$. **(F)** In situ experimental deformation of a $1 \times 3$ CI-2 (Design 2) assembly.

*Two-port module family*

Because the intra-cell material interactions are abstracted by the nonlinear spring stiffness, the impedance-guided framework can be readily extended from collinear to orthogonal two-port unit cells by redefining the control regions (Fig. 3H). Following the same procedure, we obtained collinear transmitter (CT) designs with comparably improved deformation transmission (Fig. 5A). Compared to the bar-shaped lattice (BL, Fig. 2A), for instance, CT Design 2 has increased $4.6\%$ at $\bar{N} = 1$ and $96.4\%$ at $\bar{N} = 6$ in transmission efficiency; the corresponding FEA and experimental results at $\bar{N} = 1 \sim 3$ are presented in Fig. 5B. In addition to CI and CT, the extension enables the synthesis of orthogonal inverter (OI) and orthogonal transmitter (OT) designs under the same global optimization strategy. Together, these four classes establish a family of two-port soft-metamaterial modules (Fig. 5A).

For each class, multiple geometries can be generated by constraining different total displacement gain $r_{tot}^{(i)}$, with their respective characteristic attenuation length optimized (e.g., CI-1 denotes the optimized collinear inverter by constraining $r_{tot}^{(1)}$). Owing to the tension-compression symmetry (Supplementary Fig. S7), each design admits a pair of dual transmission modes, which can be indexed by the port placement and the displacement sign in a Cartesian coordinate system. For instance, the pair $L^{(+)}R^{(+)}/L^{(-)}R^{(-)}$ corresponds to a CT design aligned along the $x$-axis. In addition, each pair of modes has coordinate variants associated with different placements of unit cell, as enumerated in Supplementary Fig. S8. Collectively, this family covers all possible one-input, one-output transmission modes within a square unit cell, thereby providing a complete description for the two-port design space.

The two-port family enables arbitrary one-dimensional (1D) assemblies with direction-changing and aperiodic cascaded layouts. We demonstrate this capability through two soft-metamaterial assemblies designed for obstacle-bypassing transmission, a task that is not readily achieved through intuitive design. The first configuration is a bent $1 \times 7$ array composed of CT and OT unit cells, designated to transmit the input displacement while preserving its direction after detouring around an obstacle. Multiple candidate assemblies can be constructed from the two-port family; one representative example combines CT-1, OT-1, and OT-2 (Figs. 5C and 5D; denoted as CT1+OT1/2), with the corresponding displacement field shown in Fig. 5E. Additional impedance-guided designs, including CT1+OT2, CT1+OT3, and CT2+OT2, are provided in Supplementary Fig. S9. Compared with the intuitively designed assembly (Fig. 5F and Supplementary Fig. S10), these optimized configurations improve the transmission efficiency by more than 9-fold, further supporting the effectiveness of directly applying the global optimization strategy to nontrivial aperiodic assemblies. We next consider a more challenging $1 \times 8$ configuration (Fig. 5G), in which the transmitted displacement reverses direction twice along the path and the output returns to a location near the input. Whereas the intuitive design exhibits almost no useful transmission, with $|D_{out}/D_{in}| < 1\%$, the impedance-guided design successfully actuates the outlet port, reaching $|D_{out}/D_{in}| > 20\%$. The detailed displacement distributions of the two assemblies are provided in Supplementary Fig. S11.

*Four-port module family*

In contrast to two-port designs, which only support one-dimensional cascaded assemblies, four-port soft-metamaterial unit cells enable two-dimensional interactions and thus offer substantially greater programmability. With four ports located at the left (L), right (R), top (T), and bottom (B) edges of a unit cell, the corresponding effective nonlinear spring model can be extendedly described by a $4 \times 4$ tangential stiffness matrix, $\boldsymbol{K} = [K_{ij}]_{4\times4}$, together with the incremental port displacement vector $\boldsymbol{\delta u} = [\delta u_L, \delta u_R, \delta u_T, \delta u_B]^T$ and force vector $\boldsymbol{\delta f} = [\delta f_L, \delta f_R, \delta f_T, \delta f_B]^T$. This formulation allows a wide range of port assignments in assembled systems, including multi-input-single-output, multi-input-multi-output, and single-input-multi-output configurations. Specifically, considering a one-input, three-output case, a family of four-port (FP) modular unit cells can be generated by applying the impedance-guided global optimization strategy (see Supplementary Text S5 for detailed formulation). Similar to the two-port family, four types of geometry (FP-1, FP-2, FP-3, and FP-4), corresponding to four pairs of dual transmission modes, provide a complete description of the four-port design space, with eight coordinate variants in total (Fig. 5H and Supplementary Fig. S12).

The ability of the established four-port family to capture assembly-level interactions is further demonstrated through both periodic and aperiodic two-dimensional assemblies. First, as shown in Fig. 5I, when single-sided inputs are applied to the top three ports of a $3 \times 3$ periodic assembly composed of FP-3 unit cells, deformation is transmitted across the structure with comparable magnitude, in accordance with the prescribed transmission directions encoded in the transmission mode. Notably, the remaining ports exhibit transmission efficiencies of $|D_{out}/D_{in}| > 17\%$, with the maximum reaching $80.08\%$ (Supplementary Fig. S13). We next consider an aperiodic assembly composed of FP-1 to FP-4, subjected to the same input conditions as the periodic configuration, with an output region of interest defined at the middle of the bottom boundary (Fig. 5J). The resulting transmission efficiencies reach $48.79\%$, $35.25\%$, and $35.22\%$ for the designated output ports, indicating the potential for grasping functionality under unilateral actuation (Supplementary Fig. S13; see below). Together, the two-port and four-port families enables effective regulation of both the direction and magnitude of the transmitted displacement and force. This capability establishes the basis for modular multifunctionality and a highly extendable design space for programmable soft metamaterials (Supplementary Movie S1).

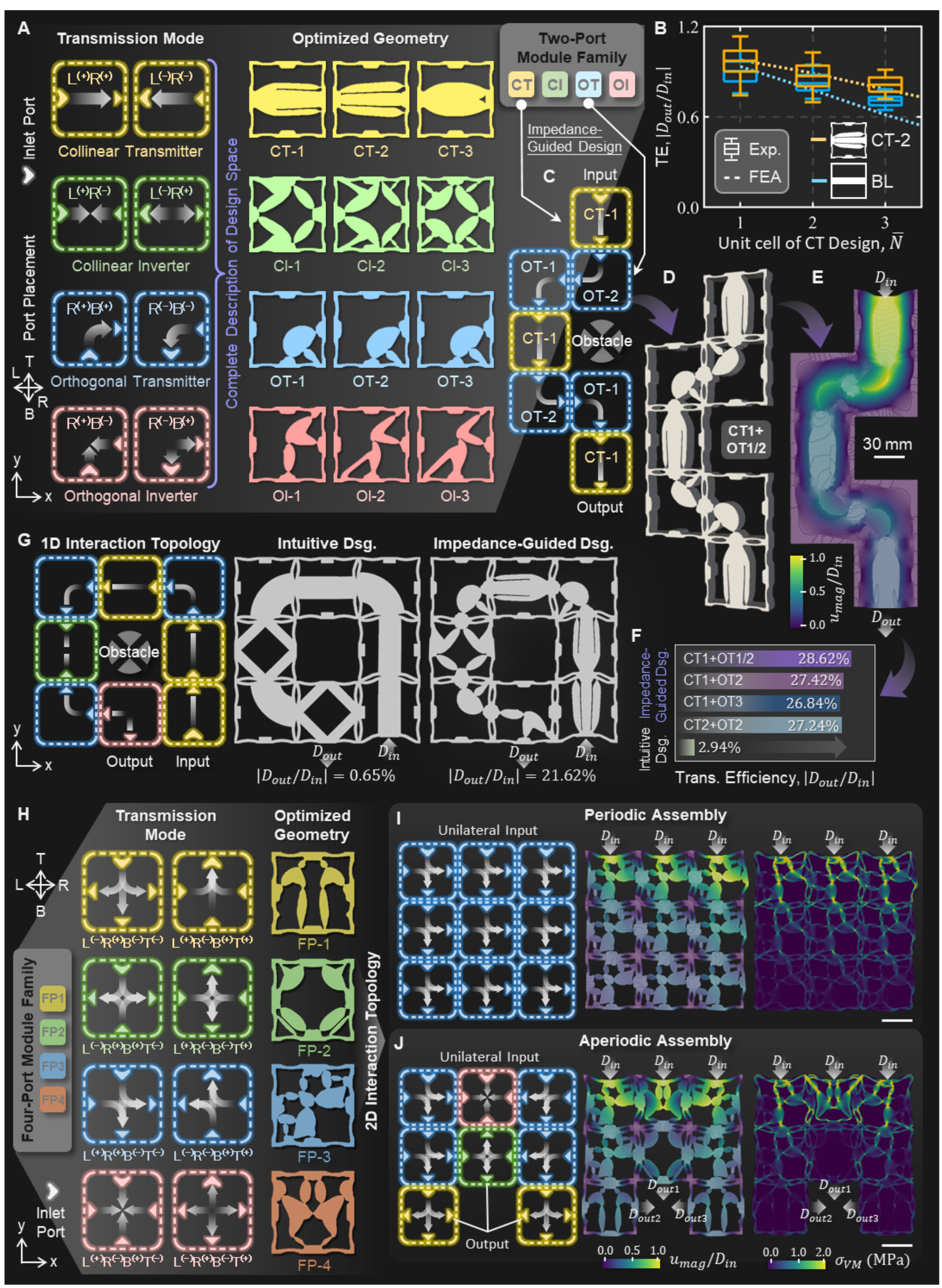

**Fig. 5. Modular families of optimized soft-metamaterial unit cells and their reconfigurable assemblies with enhanced transmission.**

**(A)** Two-port module family comprising CT, CI, OT, and OI optimized designs, together with a complete description of the transmission modes in the design space. **(B)** Comparison of transmission efficiency between the numerical and experimental results of CT-2 and the BL benchmark design at $\bar{N} = 1 - 3$. **(C)** Prescribed transmission mode of the $1 \times 7$ obstacle-bypassing assembly, denoted as CT1+OT1/2. **(D)** Geometry layout of CT1+OT1/2 assembly. **(E)** Displacement field distribution of CT1+OT1/2 under input actuation. **(F)** Comparison of transmission efficiency among the impedance-guided designs (CT1+OT1/2, CT1+OT2, CT1+OT3, and CT2+OT2) and the intuitive design. **(G)** Performance comparison of deformed configurations between the intuitive design and the impedance-guided design under the prescribed transmission mode of the $1 \times 8$ obstacle-bypassing assembly. **(H)** Four-port module family comprising FP-1, FP-2, FP-3, and FP-4, together with a complete description of the transmission modes in the design space. **(I)** Prescribed transmission mode, displacement field, and von Mises stress distribution of the periodic FP-3 assembly. **(J)** Prescribed transmission mode, displacement field, and von Mises stress distribution of the aperiodic four-port assembly.

## Programmable multifunctionalities of modular soft metamaterials

Based on the established impedance-guided design framework and the corresponding families of two-port and four-port modules, programmable multifunctionalities emerge from the modular assembly. By combining unit cells with target transmission behaviors, the soft metamaterial can be configured to realize diverse, spatially organized mechanical functions beyond simple deformation transfer.

*Single-input alternating compliant switch array*

Because the prescribed responses of each unit-cell port can be independently assigned under a given input, deformation can be distributed across a metamaterial assembly in a controllable manner to realize diverse target actuation patterns. We demonstrate this programmable deformation distribution by converting port displacement into digital signals in a light-emitting device, as illustrated in Fig. 6A. Specifically, a dual-LED module containing one blue and one red LED is attached to each port of a unit cell. Port displacement along the positive or negative axis direction closes the circuit of the blue or red LED, respectively, thereby functioning as a single-pole double-throw (SPDT) switch. These modules are then mounted sequentially along the assembly direction of the soft metamaterial, where they are triggered according to the prescribed deformation-transmission pattern and a threshold $\Delta$ (Fig. 6A; see Supplementary Fig. S14 for the detailed structural design).

Using the LED display system together with the two-port module family (Fig. 6B), we first realized an array of compliant circuit switches in which the triggered state alternates under a single input (Fig. 6C). Built on a $1 \times 3$ cascaded CI-3 assembly, the four ports are each connected to a dual-LED module. While one terminal port is actuated in compression or tension, all four modules are triggered sequentially along the transferred path. Owing to the displacement-inverting behavior of the CI unit cells, the switch states, indicated by the LED colors, alternate from one module to the next (Figs. 6D and 6E; Supplementary Movie S2). Moreover, the triggered states are fully reversed when the actuation mode switches from compression to tension. Requiring only a single actuation degree of freedom and a terminal input, this material-centric system provides an energy-efficient and reliable route to multifunctional operation.

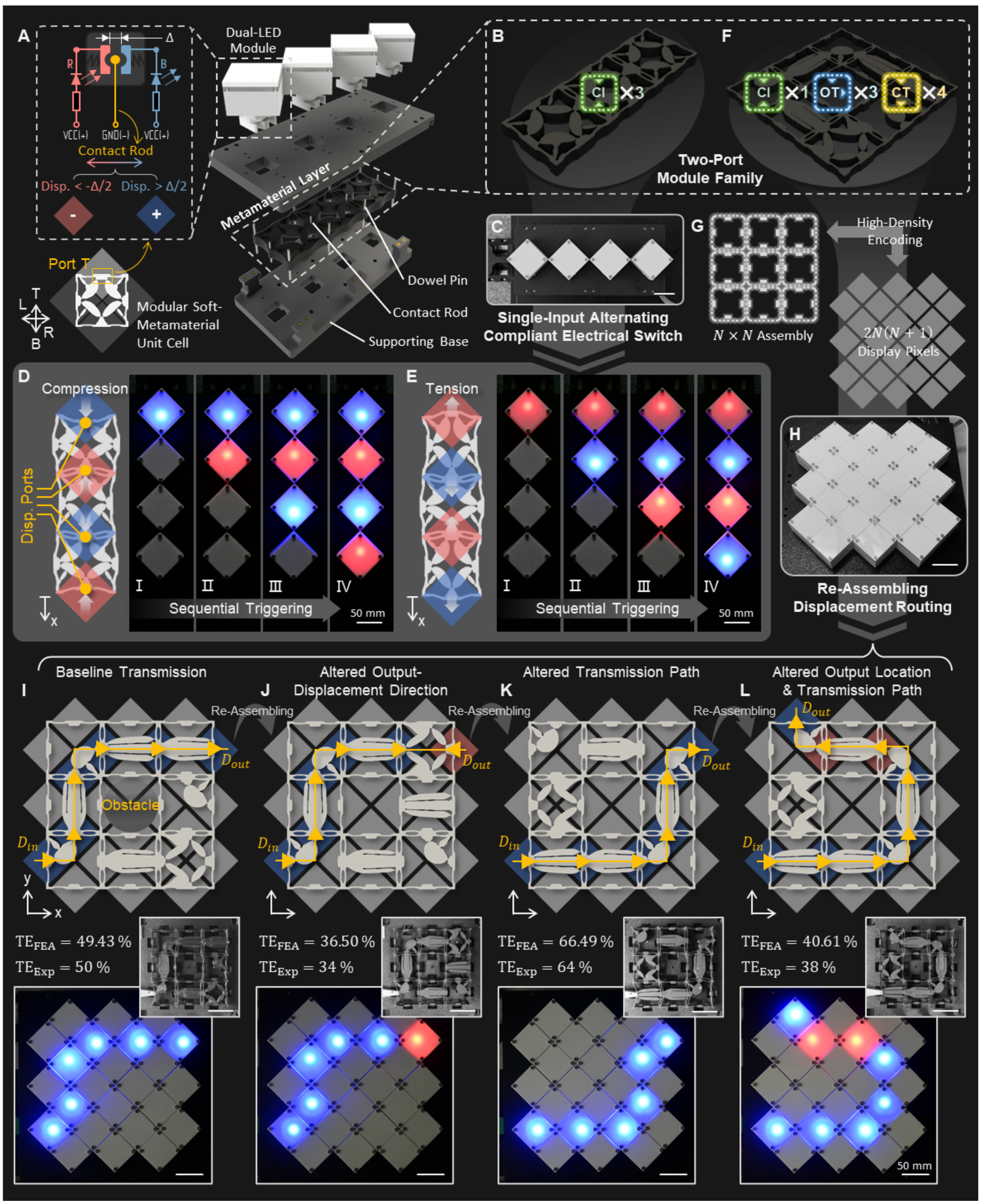

**Fig. 6. Mechanical signal routing and programmable electrical encoding through controllable deformation distribution and displacement routing.**

**(A)** Light-emitting device converting the port displacement of the metamaterial layer into the triggering of the red or blue LEDs in a dual-LED module. A contact rod embedded at each port closes the circuit once the displacement reaches $\Delta/2$, with the prescribed threshold set to $\Delta = 2$ mm. **(B)** A set of CI designs from the established two-port module family, assembled into a $1 \times 3$ soft-metamaterial structure. **(C)** The corresponding $1 \times 3$ LED display system, enabling a single-input array of alternating compliant switches. **(D)** Sequential and alternating triggering of the array under compressive actuation. **(E)** Sequential and alternating triggering of the array under tensile actuation. **(F)** A set of CI, OT, and CT designs from the two-port module family, assembled into a $3 \times 3$ soft-metamaterial structure. **(G)** High-density encoding from an $N \times N$ metamaterial assembly to $2N(N+1)$ display pixels. **(H)** The corresponding $3 \times 3$ mechanical LED display system, enabling programmable signaling through reassembled displacement-routing paths. **(I)** Baseline assembly transmitting the lower-left input to the upper-right port while bypassing the central obstacle. **(J)** Reassembled soft metamaterials with reversed output-displacement direction. **(K)** Reassembled soft metamaterials with redirected transmission path. **(L)** Reassembled soft metamaterials with relocated output port and the corresponding transmission path. The in situ deformed samples and the triggered LED display at the working-point input $D_{in} = 5$ mm are shown below in the panels of (I) to (L), respectively.

*Signal programming of re-assembled displacement routing*

We next demonstrate that the modular unit cells can be assembled and reassembled to realize different displacement-routing functions within the same set of building blocks (Fig. 6F). Meanwhile, the routed port displacements can be encoded as binary signals using the designed light-emitting device. A notable advantage of this mechanical LED display is its high information density: an $N \times N$ assembly can encode $2N(N+1)$ independently defined display pixels. Accordingly, a $3 \times 3$ assembly supports 24 diagonal display pixels (Fig. 6H; see Supplementary Fig. S15 for the detailed structural design).

The baseline assembly, composed of four CT, one CI, and three OT unit cells (Fig. 6F), transmits displacement through five soft unit cells from the lower-left port to the upper-right port, bypassing the central obstacle with only half attenuation (Fig. 6I). The transferred path is simultaneously visualized by the $3 \times 3$ LED display, where the displacement direction at each port is indicated by the LED color. After reassembly, the same group of unit cells can realize distinct routing behaviors: the output direction can be reversed while preserving the same transferred path (Fig. 6J), the transferred path can be altered while retaining the same output port (Fig. 6K), or the target output port can be reassigned while maintaining considerable transmission efficiency (Fig. 6L). A detailed actuation sequence of the displacement-routing assembly is provided in Supplementary Movie S3.

*Delicate grasping of modular soft-metamaterial gripper*

The high extendibility and large design freedom of the module assemblies are further demonstrated by integrating the numerical four-port modular design in Fig. 5J into a physical soft robotic gripper. Specifically, we designed a layered structure by mounting the modular assembly between a pair of fixed brackets, as illustrated in Fig. 7A (see Supplementary Fig. S16 for detailed structure design). With a one-unit-cell-long grasping region prescribed, adjacent ports of the assembly serve as the contact ends of the gripper and actuate according to the designated transmission behavior (Supplementary Fig. S13).

The grasping performance of the modular soft gripper is first evaluated through load-lifting tests (Fig. 7B and Supplementary Movie S4). By varying the size and weight of cubic grasped object while keeping the surface condition and input working point fixed as $D_{in} = 5$ mm, the gripper exhibits reliable lifting over a one-side stroke of $1.75$ mm ($3.50$ mm in total), reaching a maximum lifting efficiency of $59.57$ g/mm. More importantly, its delicate-handling capability is highlighted by a feasible grasping window for fresh chicken eggs, used here as representative fragile objects (Fig. 7C). With masses of approximately 60-70 g, eggs of various sizes were successfully gripped over multiple repeated trials using an open-looped actuator. By contrast, under the same $D_{in}$, the intuitively designed soft-metamaterial assembly generated an excessive load at the gripping port that exceeded the fracture threshold of the eggs, causing shell cracking across all tested egg sizes (Fig. 7D). The improved grasping of fragile objects can be attributed to the higher transmission efficiency of the modular-cell architecture, which reduces parasitic energy storage and mitigates excessive contact force under the same input displacement.

We further demonstrate the strength of modularity by introducing a local defect into the soft gripper. While one unit cell near the single-sided input was removed, the egg-grasping performance was barely affected (Fig. 7E; Supplementary Movie S4), indicating local-defect tolerance enabled by the enhanced 2D interactions of the four-port metamaterial assembly. Beyond the symmetric $3 \times 3$ gripper presented here, the impedance-guided design framework can be readily extended to a wide range of modular assemblies, enabling diverse soft actuators with strong generalizability and robustness.

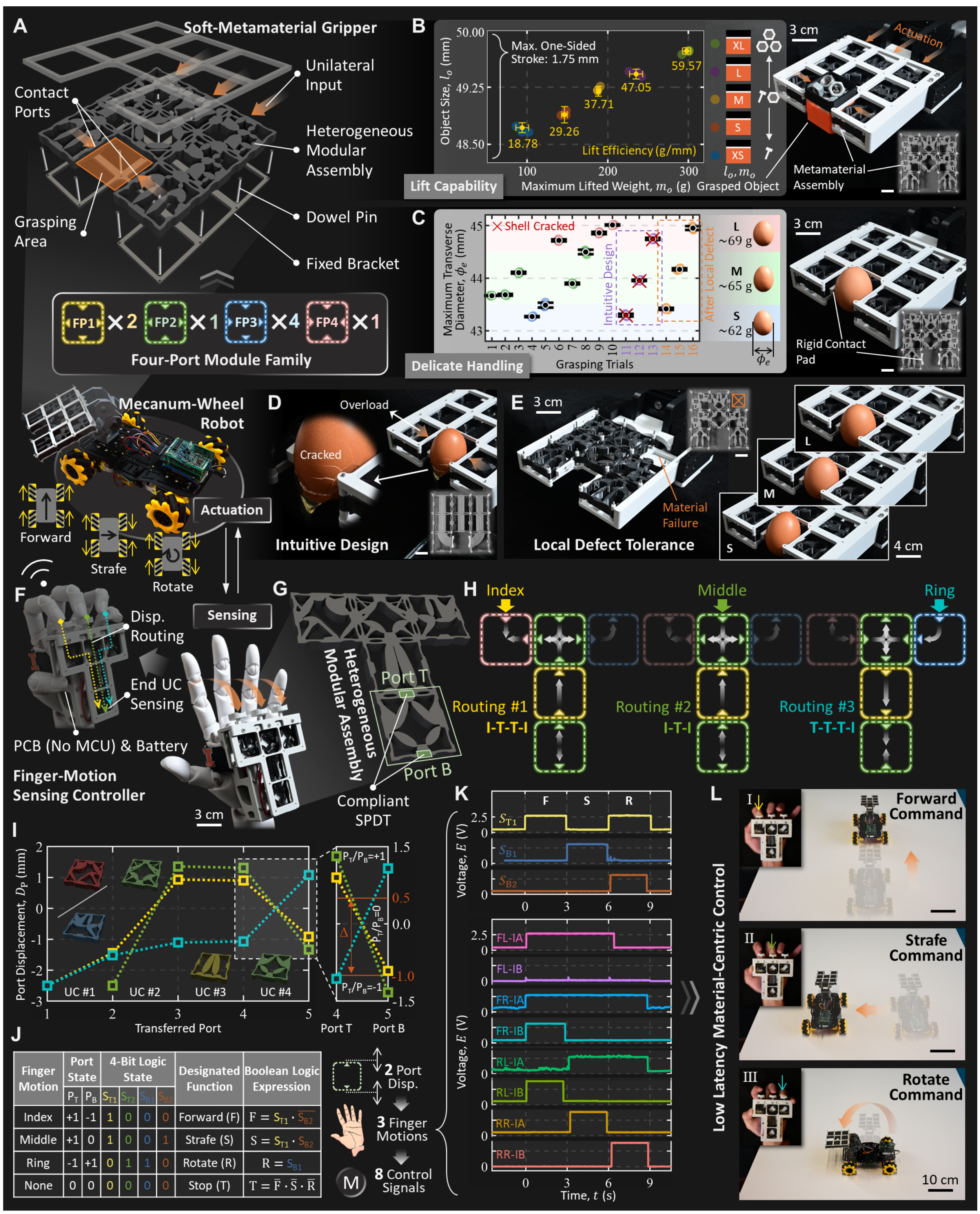

**Fig. 7. Programmable soft sensing and actuation embedded in modular soft-metamaterial assemblies.** **(A)** Soft-metamaterial gripper based on a heterogeneous modular assembly from the four-port module family. **(B)** Load-lifting test, quantified by the maximum lifted weight. The grasped objects were cubic PLA boxes of different sizes, and the load was increased gradually using standard nuts and screws until the object slipped or fell. Repeated measurements of object size and maximum lifted weight were performed six and three times,

respectively. Colored circular markers denote individual data points, and yellow error bars indicate standard deviations. The actuated experimental setup and the integrated TPU soft-metamaterial sample are shown on the right. **(C)** Delicate-handling test based on repeated grasping trials using commercially available unfertilized chicken eggs as fragile objects. Repeated measurements of the egg transverse diameter were performed three times. Colored circular markers denote individual data points, and black error bars indicate standard deviations. The actuated experimental setup and the embedded TPU soft-metamaterial sample are shown on the right. **(D)** Egg-gripping experiment using the intuitively designed assembly, showing shell cracking during grasping. **(E)** Local-defect-tolerance test performed by removing one unit cell near the actuation end, while retaining egg-grasping functionality across all tested object sizes. **(F)** Demonstration of a finger-motion-sensing controller used to remotely operate a four-motor Mecanum-wheel robot. **(G)** Optimized geometry of the heterogeneous modular assembly integrated into the controller. **(H)** Three displacement-routing paths corresponding to the index, middle, and ring fingers, constructed from different combinations of inverter and transmitter modules. **(I)** Variation of the transferred port displacement under the three finger inputs, together with a zoomed-in view of the end unit cell. A prescribed SPDT threshold $\Delta = [-1.0, +0.5]$ mm is defined accordingly to distinguish the states of ports T and B. **(J)** Logic-function mapping table relating the states of the two ports to three designated robot functions. **(K)** Measured voltage of the port logic states and the corresponding wheel-motor driver signal channels during a sequential Forward-Strafe-Rotate (F-S-R) command sequence. **(L)** Remote-control experiment showing low-latency robot motion driven by transmitted finger-motion-sensing signals.

*Finger-motion sensing and remote-control controller*

Finally, we developed a wearable finger-motion-sensing controller that detects finger compression and remotely operates a Mecanum-wheel robot (Fig. 7F). In this system, the designed soft-metamaterial assembly serves as a mechanical transduction layer, transferring localized inputs from the fingertips to the end unit cell near the proximal palmar region (Fig. 7F). By converting the transmitted deformations of the last two ports (ports T and B) into the closing states of compliant SPDT switches (Fig. 7G), the port displacements can be mapped into logic states. As a result, inputs from the index, middle, and ring fingers, each associated with a distinct transmission path and length, encode different robot-motion commands (Fig. 7F). Notably, no electrical wiring is required throughout the assembly except at the terminal unit cell, yielding a material-centric sensing device (see Supplementary Fig. S17 for detailed structure design). It is also advantageous for hazardous operations where close access is not feasible.

The transferred displacement-routing paths corresponding to the three finger inputs are shown in Fig. 7H. In this design case, we adopted the proposed recursive optimization strategy to ensure effective displacement routing (Supplementary Fig. S4). Starting from the end unit cell, we sequentially optimized the geometries of the five target unit cells. The resulting displacement fields (Supplementary Fig. S18) and the transferred port displacements (Fig. 7I) show effective transmission with distinct responses to different inputs. Accordingly, different port-state combinations can be defined by prescribing a triggering threshold $\Delta$. On this basis, we encoded each finger-associated function into the logic state of the two ports (Fig. 7J), and further converted the forward (F), strafe (S), and rotate (R) commands into eight control signals for the four wheel-motor drivers using a simple logic circuit (Supplementary Fig. S19; Supplementary Table S4). Using the metamaterial-processed sensing signals, we achieved remote control of the robot without a microcontroller (Figs. 7K and 7L; Supplementary Movie S5), yielding a compact and low-latency wearable controller. This result highlights the versatility of the impedance-guided design framework under a stringent localized-input condition, extending deformation programming beyond displacement-field control toward flexible material-based information processing.

## Conclusions

This work establishes a mechanics framework for understanding and programming localized deformation transmission in modular soft metamaterials. By focusing on the essential mechanical behavior governing attenuation of localized deformation in soft metamaterials and introducing an equivalent nonlinear spring representation, we quantify the intra- and inter-unit-cell interactions and derive an elastostatic input-output relation that captures the assembly-level behavior using only unit-cell-level analysis. Building on this formulation, we developed an impedance-guided generative design framework that markedly enhances transmission efficiency in modular soft assemblies, whereas existing intuitive design strategies and conventional compliant mechanism optimization approaches show limited capability for this class of problems. The optimization histories further reveal that reducing transmission attenuation drives the effective mechanical properties toward extreme regimes, accompanied by an emergent mechanized morphology that suppresses strain-energy storage. Accounting for the intrinsic nonlinearity of soft metamaterials, we established two-port and four-port module families that provide a complete description of the corresponding two-dimensional interaction and deformation modes. Leveraging the modularity of these designs, we demonstrated their nontrivial and adaptable programmability through alternating compliant switches, displacement-routing assemblies, and a local-defect-tolerant soft gripper. beyond mechanical actuation, the programmable deformation transmission in the modular assemblies enables information-processing capabilities in an energy-efficient and low-latency manner. They not only encode diverse transmission states in a mechanical LED display, but also function as finger-motion-sensing metamaterial controllers, in which the input deformation is transmitted to the output end while electrical integration is required only at a single unit cell. Together, these results highlight the potential of the impedance-guided framework for handling complex assembly-level interactions and establish a versatile route toward large-scale material systems in which modularity and scalability are essential, thereby advancing the development of embodied intelligent metamaterials. Future work will explore more unconventional unit-cell responses, including multistability and nonreciprocity, as well as broader integration with multiphysics coupling to further expand functionality.

## Materials and Methods

### Nonlinear finite element analysis (FEA) with self-contact

To capture the nonlinear behavior of the soft material, we adopt the first-invariant ($I_1$) based hyperelastic model proposed by Lopez-Pamies [38], which provides a more accurate constitutive description for rubber-like elastomers than the Neo-Hookean model. Self-contact during large deformation is further incorporated using the third-medium contact method [39,40]. The resulting nonlinear finite element problem is formulated in residual form and solved using the Newton-Raphson method. All finite element calculations were carried out in FEniCSx 0.7.2 [41] with PETSc-based solvers [42]. Further details are provided in Supplementary Text S6 and Supplementary Fig. S20.

**Topology optimization**

Following the solid isotropic material with penalization (SIMP) method [43], the design domain $\Omega$ of size $L \times L$ was discretized using a $200 \times 200$ structured quadrilateral mesh, with the design variable defined as material density $\boldsymbol{\rho} = \{\bar{\rho}_e\}\ (e = 1,\ \ldots\ , N_e)$ (Fig. 2C and 3H). A boundary condition was adopted by fixing the four corners of the unit cell, which facilitates the scalability and modularity of the soft-metamaterial assembly. A thin-walled solid passive zone was further prescribed near the edges to define a clear unit-cell connection boundary (Supplementary Fig. S22) and to ensure that deformation was transmitted only through the port regions (controlled regions). Sensitivities of the objective and constraints were evaluated using the adjoint method, and the design variables were updated using the method of moving asymptotes (MMA) [44]. Further details are provided in Supplementary Texts S8 and S9 and Supplementary Figs. S23 to S25.

**Quantification of the effective resistance perceived at the port of unit cell**

With the equivalent spring stiffness matrix $\boldsymbol{K}$ (Supplementary Text S1), the port perturbations can be mapped to the port reactions $(\delta u_{in}, \delta u_{out}) \mapsto (\delta f_{in}, \delta f_{out})|_{u_{in}=D_{in}}$ as:

$$\begin{bmatrix} \delta f_{in} \\ \delta f_{out} \end{bmatrix} = \begin{bmatrix} K_{11} & K_{12} \\ K_{21} & K_{22} + k_{out} \end{bmatrix} \begin{bmatrix} \delta u_{in} \\ \delta u_{out} \end{bmatrix}. \quad (4)$$

In a passive and reciprocal system, we can have the positive definition as $K_{11} > 0$, $K_{22} + k_{out} > 0$, and the determinant $\det \boldsymbol{K} = K_{11}(K_{22} + k_{out}) - K_{21}^2 > 0$. According to Eq. 4, the incremental transmission ratio for a displacement perturbation can be defined as

$$r(k_{out}) := \frac{\delta u_{out}}{\delta u_{in}} = -\frac{K_{21}}{K_{22} + k_{out}}. \quad (5)$$

which characterizes the local transmission behavior at a specific working point. Its sign determines the direction of the incremental output response: $K_{21} > 0$ corresponds to an inverting response, whereas $K_{21} < 0$ corresponds to a non-inverting one. The corresponding incremental input stiffness of the unit cell is

$$K_{in}(k_{out}) := \frac{\delta f_{in}}{\delta u_{in}} = K_{11} + K_{12} r(k_{out}) = K_{11} - \frac{K_{21}^2}{K_{22} + k_{out}}, \quad (6)$$

which quantifies the effective resistance perceived at port L under the output constraint imposed by $k_{out}$. For an $\bar{N} \times 1$ array of two-port unit cells assuming a lossless connection between adjacent unit cells, the interface conditions are given by displacement continuity, $u_{in}^{(i+1)} = u_{out}^{(i)}$, and force equilibrium, $f_{in}^{(i+1)} + f_{out}^{(i)} = 0$. Accordingly, the $i$-th unit cell can be regarded as being attached to an equivalent nonlinear spring with stiffness $k_{out}^{(i)}$, which represents the collective elastic resistance of cells $i+1$ to $\bar{N}$. The incremental input stiffness of cell $i$+1 will serve as the external resistance at the output of its upper-level unit cell ($i$-th), written as

$$k_{out}^{(i)} = K_{in}^{(i+1)}\left(k_{out}^{(i+1)}\right). \quad (7)$$

In this case, the modular transmission ratio and the input impedance for the $i$-th unit can be expressed as

$$r^{(i)} = -\frac{K_{21}^{(i)}}{K_{22}^{(i)} + Z_{in}^{(i+1)}}, \quad (8)$$

and

$$Z_{in}^{(i)} = K_{11}^{(i)} - \frac{\left(K_{21}^{(i)}\right)^2}{K_{22}^{(i)} + Z_{in}^{(i+1)}}. \quad (9)$$

Eq. 9 is a recursive formula of Riccati-form, indicating the deformation transmission's dependence on the length of the cascaded array. Referring to Eq. 8, the total displacement gain from the first input to the $\bar{N}$-th output can be then defined as

$$r_{tot}^{(\bar{N})} = \frac{\delta u_{out}^{(\bar{N})}}{\delta u_{in}^{(1)}} = \prod_{i=1}^{\bar{N}} r^{(i)} = \prod_{i=1}^{\bar{N}}\left(-\frac{K_{21}^{(i)}}{K_{22}^{(i)} + Z_{in}^{(i+1)}}\right), \quad (10)$$

where $\mathrm{sign}\left(r^{(i)}\right) = -\,\mathrm{sign}\left(K_{21}^{(i)}\right)$ determines the direction of the transferred displacement at each cascaded level.

**3D-printing of TPU and PLA**

TPU was selected as the matrix material for the designed soft metamaterials because of its high elasticity and toughness. Samples were fabricated using an FDM 3D printer (Bambu Lab H1D, China), with TPU filament (durometer 85A; tensile strength, 580 psi) printed at $235\ ^\circ\mathrm{C}$ on a $65\ ^\circ\mathrm{C}$ PEI build plate. A constant thickness of $10\ \mathrm{mm}$ was assigned to all optimized two-dimensional designs. PLA was used for the rigid frames and structural components, with a nozzle temperature of $220\ ^\circ\mathrm{C}$ and a bed temperature of $60\ ^\circ\mathrm{C}$. The printing speed was set to $15\ \mathrm{mm/s}$ for all parts.

**Material characterization of TPU**

To characterize the mechanical properties of TPU, tensile specimens (ASTM D638 Type IV) were printed and tested at a stretching rate of $5\ \mathrm{mm/min}$ using a universal testing machine (Instron 6800, UK). The displacement and strain fields were measured by digital image correlation (DIC) [45]. The resulting stress-stretch curves were then fitted by the adopted LP model for the FEA simulation. Further details are provided in Supplementary Text S7 and Supplementary Fig. S21.

**Input actuation for soft-metamaterial assembly**

In the experiments, linear displacement input was provided by a linear-stage actuator (Yeebyee SFU1605, China) mounted on a T-slotted aluminum frame with 3D-printed adapter components. The actuator was driven by an integrated NEMA 17 stepper motor (1.8° step angle, 1.5 A, and 1.2 N·m holding torque) through a TB6600 driver at 16x microstepping, yielding a linear resolution of 640 steps/mm. The stepper motor was controlled by an Arduino Uno through serial commands, with the displacement speed set to approximately 1.04 mm/s. During actuation, an input displacement of $D_{in} = 5$ mm was applied to the input port of each assembly, followed by retraction to the home position.

**Port displacement measurement of soft-metamaterial assembly**

Samples of the CI-1, CI-2, CI-3, DS, CT-2, and BL designs, assembled with $\bar{N} = 1$ to $3$ unit cells, were 3D-printed and tested to measure the transferred output displacement under a localized input of $D_{in} = 5$ mm. an image-processing pipeline was used to achieve pixel-level measurement accuracy, including image calibration, rigid-body-motion correction, and edge detection. Further details are provided in Supplementary Text S10 and Supplementary Figs. S26 and S27.

## CRediT Authorship Contribution Statement

**W.X.**: Conceptualization, Investigation, Methodology, Visualization, Writing- original draft preparation. **D.H.**: Data curation, Validation, Visualization. **Z.Z.**: Investigation, Writing-reviewing and editing. **R.D.K.**: Validation, Writing- reviewing and editing. **X.S.Z.**: Conceptualization, Investigation, Writing- reviewing and editing, Supervision.

## Conflict of Interests

The authors declare no competing interests.

## Acknowledgments

The authors acknowledge support from the Air Force Office of Scientific Research (award FA9550-23-1-0297) and National Science Foundation (grants CMMI-2344258 and CMMI-2505649). The authors are grateful to Mr. Bharath Iyer for helpful suggestions regarding the remote-control scheme used in the finger-motion-sensing controller.

## Data Availability

The data and code supporting the findings of this study are available from the corresponding authors upon reasonable request.